\documentclass[11pt,a4paper]{amsart}
\usepackage{epsfig}
\usepackage{amsmath, amssymb}
\usepackage[latin1]{inputenc}
\usepackage[english]{babel}
\usepackage{graphicx}
\usepackage{graphics}
\usepackage{multirow}
\usepackage{wrapfig}
\usepackage{hyperref}
\hypersetup{colorlinks=true,linktoc=page,linkcolor=blue,citecolor=red,urlcolor=blue}

\title{Semi-transparent Boundary Conditions in the Worldline Formalism}

\author{S.A.\ Franchino Vi\~nas}%
\address{IFLP - CONICET / Departamento de F\'{\i}sica, Facultad de Ciencias Exactas, Universidad Nacional de La Plata, C.C.\ 67, (1900)  La Plata, Argentina}%
\email{safranchino@yahoo.com.ar}%

\author{P.A.G.\ Pisani}%
\address{IFLP - CONICET / Departamento de F\'{\i}sica, Facultad de Ciencias Exactas, Universidad Nacional de La Plata, C.C.\ 67, (1900)  La Plata, Argentina}%
\email{pisani@fisica.unlp.edu.ar}%

\date{\today}

\begin{document}

\maketitle
\begin{abstract}
The interaction of a quantum field with a background containing a Dirac delta function with support on a surface of codimension 1 represents a particular kind of matching conditions on that surface for the field. In this article we show that the worldline formalism can be applied to this model. We obtain the asymptotic expansion of the heat-kernel corresponding to a scalar field on $\mathbb{R}^{d+1}$ in the presence of an arbitrary regular potential and subject to this kind of matching conditions on a flat surface. We also consider two such surfaces and compute their Casimir attraction due to the vacuum fluctuations of a massive scalar field weakly coupled to the corresponding Dirac deltas.

\end{abstract}

\section{Introduction}

The Worldline Formalism (WF) is an intuitive and efficient method to compute effective actions in Quantum Field Theory (QFT) \cite{Schubert:2001he,Bastianelli:2006rx}. This formalism has been applied to the coupling of external gravity to quantum fields of spin 0 \cite{Bastianelli:2002fv}, spin 1/2 \cite{Bastianelli:2002qw}, spin 1 and, more generally, antisymmetric tensor fields \cite{Bastianelli:2005vk}. Higher spin fields have also been studied in the WF for flat space \cite{Bastianelli:2007pv} and a conformally flat background \cite{Bastianelli:2008nm}. The method has also proved useful for the study at the quantum level of photon-graviton mixing in an electromagnetic background \cite{Bastianelli:2004zp}.

In the last years, the first steps of a systematic generalization of the WF to manifolds with boundaries have been taken through the computation of the asymptotic expansion of the heat-kernel trace for a scalar field on a flat manifold with boundary \cite{Bastianelli:2006hq,Bastianelli:2007jr,Bastianelli:2008vh,Bastianelli:2009mw}.

Worldline technics combined with Monte Carlo methods have also been successfully applied to compute Casimir energies for various geometries \cite{Gies:2003cv,Gies:2006bt,Gies:2006cq}, giving results on the edge effects \cite{Gies:2006xe} as well as an interesting behavior of the Casimir force in terms of the temperature \cite{Gies:2008zz,Weber:2010kc,Weber:2010xv}. One of the advantages of worldline Monte Carlo with respect to numerical calculation on the lattice is that the former can be implemented on a fermionic system without a discretization of spacetime \cite{Dunne:2009zz}.

In this article we show that the WF can be also applied to a field under the matching conditions on a surface of codimension 1 imposed by a background containing a Dirac delta with support on that surface. To do that, we obtain the asymptotic expansion of the heat-kernel corresponding to a scalar field on $\mathbb{R}^{d+1}$ interacting with a delta function with support on a flat $d$-dimensional surface. Later, in the same formulation, we study the Casimir force between two such surfaces due to the vacuum oscillations of the scalar field.

A Dirac delta potential with support on a surface of codimension 1 is a well-defined mathematical problem \cite{Albe} which has been extensively studied to model several physical configurations.

In the last years, this setting has been applied to quantize the TE-modes of the electromagnetic field in the vicinity of systems such as giant carbon molecules \cite{Barton1,Barton2,Barton3,Bordag:2005qv,Bordag:2005by,Bordag:2006kx,Bordag:2007zz,Bordag:2008rc,Bordag:2009zzc}. The collective electrons of these molecules can be considered as an infinitely thin plasma layer that imposes semi-transparent boundary conditions on the oscillation modes of the quantum electromagnetic field. For small values of the net charge and current at the layer the problem reduces -for TE-modes- to solving the wave equation with a delta-like potential at the layer.

Effective actions for quantum fields at the 1-loop level can be computed using zeta-function and heat-kernel technics \cite{Kirsten:2001wz,Vassilevich:2003xt}. In particular, the asymptotic expansion of the heat-kernel for small values of the proper time contain information of the 1-loop ultraviolet divergencies. In the WF this asymptotic expansion corresponds to the semiclassical expansion of a field theory in 0+1 dimensions.

Let us consider a scalar field $\phi(x)$ defined on $x=(x_1,\ldots,x_{d+1})\in\mathbb{R}^{d+1}$ with Euclidean action
\begin{align}\label{act}
    S[\phi]=\int_{\mathbb{R}^{d+1}}\left[\frac{1}{2}\,\partial_i \phi(x)\partial_i\phi(x)+U(\phi(x))\right].
\end{align}
The corresponding 1-loop effective action is given by
\begin{align}\label{effact}
    \Gamma[\phi]=S[\phi]+\frac{1}{2}\log{\rm Det}A\ ,
\end{align}
where the operator of quantum fluctuations $A$ -for the action (\ref{act})- is the differential operator
\begin{align}\label{schope}
    A=
    -\triangle+V(x)
    :=-\triangle+U''(\phi(x))\ .
\end{align}
The primes on $U(\phi)$ denote derivatives with respect to $\phi$. The functional determinant of the Schr\"odinger operator $A$ can be defined by means of its corresponding heat-kernel $K$. A usual definition for the functional determinant in expression (\ref{effact}) is \cite{Vassilevich:2003xt}
\begin{align}\label{dethea}
    \log{\rm Det}\,A:=-\int_0^\infty \frac{d\beta}{\beta}\ {\rm Tr}\,e^{-\beta A}\ ,
\end{align}
where
\begin{align}\label{tra}
    {\rm Tr}\,e^{-\beta A}=\int_{\mathbb{R}^{d+1}}K(x,x,\beta)\
\end{align}
and the heat-kernel $K$ is defined as the solution of the system
\begin{align}\label{hkequ}
\begin{split}
\left(\partial_\beta+A\right)K(x,x',\beta)&=0\ , \qquad
K(x,x',0)=\delta(x,x')\ .
\end{split}
\end{align}

The divergencies at $\beta=0$ of the integral in eq.\ (\ref{dethea}) are related to the ultraviolet divergencies of the effective action. Consequently, the regularization of these divergencies can be carried out by performing the integral in $\beta$ from $\Lambda^{-1}$ to $\infty$, where $\Lambda$ plays the role of an ultraviolet cut-off regulator.

For a differential operator of the type (\ref{schope}) defined on a compact $(d+1)$-dimensional base manifold, it is a well-known result of Spectral Geometry \cite{Gilkey} that -under quite general hypothesis on the manifold, the boundary conditions and the regularity of the potential- the heat-kernel trace ${\rm Tr}\,e^{-\beta A}$ admits the following asymptotic expansion for small values of the proper time $\beta$
\begin{align}\label{traexp}
    {\rm Tr}\,e^{-\beta A}\sim
    \frac{1}{(4\pi\beta)^{(d+1)/2}}\ \sum_{n=0}^\infty
    b_{n/2}\,\beta^{n/2}\ ,
\end{align}
where the Seeley-de Witt coefficients $b_{n/2}$ are integrals on the base manifold and its boundary of local invariants built from the potential $V$ and the geometry of the base manifold. If, as already mentioned, one uses a cut-off regularization of the integral in eq.\ (\ref{dethea}) then, as can be seen from eq.\ (\ref{effact}), the first coefficients $b_{n/2}$ give the 1-loop counterterms that regularize the effective action. It is important to mention that the coefficients $b_{n/2}$ for odd $n$, i.e. those multiplying half-integer powers of $\beta$, vanish if the base manifold has no boundary.

A different asymptotic expansion of the heat-kernel could be obtained if one relaxes some of the regularity hypothesis required for the validity of (\ref{traexp}); many examples along this line have been extensively studied (see the review \cite{Vassilevich:2003xt}). In particular, the asymptotic expansion of the heat-kernel trace for a Schr\"odinger operator whose potential contains a Dirac delta function has been determined, up to our knowledge, for the first time in \cite{Bordag:1999ed}.

The purpose of this article is to show that the worldline formalism can be used as a very efficient tool to study this kind of singular potentials. In Section \ref{regpot} we will review some basics of the WF by computing the asymptotic expansion of the heat-kernel $K(x,x',\beta)$ for small $\beta$ corresponding to a Schr\"odinger operator $A$ given by (\ref{schope}) with an arbitrary regular potential $V(x)$. We will then replace this expansion into expression (\ref{tra}) to obtain, according to expansion (\ref{traexp}), the first Seeley-de Witt coefficients. In Section \ref{delpot} we will generalize the WF for the case in which the potential is not regular but contains a Dirac delta with support on a flat layer. Finally, in Section \ref{casfor} we use these techniques to compute the Casimir energy between two such layers in the weak coupling limit.

\section{Regular potential}\label{regpot}

In this section we will show how the WF works by computing the first terms in the asymptotic expansion of the heat-kernel corresponding to the Schr\"odinger operator (\ref{schope}). Moreover, in the next section this result will prove useful to extend the WF to the case where the potential contains terms with Dirac deltas with support on flat surfaces.

First of all, notice that according to eqs.\ (\ref{hkequ}), the heat-kernel $K(x,x',\beta)$ is the wave function at $x$ of a particle that was initially at $x'$ and has evolved an Euclidean time $\beta$ with Hamiltonian $A=-\triangle+V(x)$. This is the first step in the implementation of the WF: since the operator of quantum fluctuations $A$ that determines the 1-loop effective action (see eq.\ (\ref{effact})) depends on the field theory under consideration, one must find for each field theory an auxiliary quantum mechanical particle with Hamiltonian $A$, such that the corresponding heat-kernel could be regarded as its wave function.

In the case of the self-interacting scalar field with action (\ref{act}) the operator $A$ is given by (\ref{schope}) and therefore corresponds to the Hamiltonian of a non-relativistic Schr\"odinger particle. The next step in the WF consists in implementing the path integral quantization of this auxiliary particle:
\begin{align}\label{eq.2.wlf}
{K}(x,x',\beta)&=
\int\limits_{\substack{x(0)=x'\\x(\beta)=x}} \mathcal{D}x(\tau)
\ e^{\textstyle-\int^{\beta}_0 dt\,\left[\dot{x}^2(t)/\!4+V\left(x(t)\right)\right]}\ .
\end{align}
In order to get an expansion for small $\beta$, it is convenient to make the following rescaling in Euclidean time $\tau\rightarrow \beta \tau$ so that
\begin{align}\label{eq.2.wlf2}
{K}(x,x',\beta)&=
\int\limits_{\substack{x(0)=x'\\x(1)=x}} \mathcal{D}x(\tau)
\ e^{\textstyle-\int^{1}_0 dt\,\left[\dot{x}^2(t)/\!4\beta+\beta\, V\left(x(t)\right)\right]}\ .
\end{align}
This expression is indeed appropriate for a small $\beta$-expansion for two reasons. On the one hand, since $\beta$ acts as the intensity of the potential term, one gets a series in powers of $V$
\begin{align}
\label{eq.2.exponencial.desarrollada}
{K}(x,x',\beta)&=\sum_{n=0}^{\infty}\frac{(-\beta)^n}{n!}\int\limits_{\substack{x(0)=x'\\
x(1)=x}} \mathcal{D}x(\tau)
\;e^{\textstyle-\int^{1}_0 dt'\,\dot{x}^2(t')/\!4\beta}\; \left[\int^1_0 dt\, V\left(x(t)\right)\right]^n\ .
\end{align}

On the other hand, the kinetic term shows that the propagator is proportional to $\beta$, so a complete expansion for small $\beta$ can be finally obtained by a loop expansion in the sense of the $0+1$ quantum field $x(\tau)$. In order to do that we consider the quantum field $x(\tau)$ in terms of the fluctuations $\varphi(\tau)$ about the classical trajectory with respect to the case $V\equiv 0$
\begin{align}\label{clastraj}
    x_i(\tau)=(x_i-x_i')\,\tau+x'_i+\varphi_i(\tau)\ .
\end{align}
Notice that the perturbation fields $\varphi(\tau)$ satisfy Dirichlet boundary conditions at $\tau=0$ and $\tau=1$. Replacing these new fields $\varphi(\tau)$ in eq.\ (\ref{eq.2.exponencial.desarrollada}) straightforwardly gives
\begin{align}
\label{eq.2.exponencial.desarrollada2}
\nonumber{K}(x,x',\beta)&=e^{\textstyle-\frac{(x-x')^2}{4\beta}}\sum_{n=0}^{\infty}\frac{(-\beta)^n}{n!}
\int\limits_{\substack{\varphi(0)=0\\
\varphi(1)=0}} \mathcal{D}\varphi(\tau)
\;e^{\textstyle-\int^{1}_0 dt'\,\dot{\varphi}^2(t')/\!4\beta}\;\times
\\
\nonumber
&\hspace{5cm}\times\left[\int^1_0 dt\, V\left((x-x')\,t+x'+\varphi(t)\right)\right]^n\\\nonumber&\\
&=e^{\textstyle-\frac{(x-x')^2}{4\beta}}\sum_{n=0}^{\infty}\frac{(-\beta)^n}{n!}
\left\langle\left(\int^1_0 dt\, V\left((x-x')\,t+x'+\varphi(t)\right)\right)^n\right\rangle\ .
\end{align}
In the last equation, we have defined the mean value
\begin{equation}
\langle\ \ldots\ \rangle:=\int\limits_{\substack{\varphi(0)=0\\
\varphi(1)=0}} \mathcal{D}\varphi(\tau)
\;e^{\textstyle-\int^{1}_0 dt'\,\dot{\varphi}^2(t')/\!4\beta}\ \ldots\ ,
\end{equation}
where the dots represent a function of the fields $\varphi(\tau)$.

At this point, it is important to remark that expression (\ref{eq.2.exponencial.desarrollada2}) coincides with the free heat-kernel ${K}_0(x,x',\beta)$ corresponding to $V\equiv 0$
\begin{align}
    {K}_0(x,x',\beta)=\frac{e^{\textstyle-\frac{(x-x')^2}{4\beta}}}{(4\pi\beta)^{(d+1)/2}}\ ,
\end{align}
if we normalize
\begin{align}\label{norm}
    \langle1\rangle:=\int\limits_{\substack{\varphi(0)=0\\
\varphi(1)=0}} \mathcal{D}\varphi(\tau)
\;e^{\textstyle-\int^{1}_0 dt'\,\dot{\varphi}^2(t')/\!4\beta}:=
\frac{1}{(4\pi\beta)^{(d+1)/2}}\ .
\end{align}

The next step is to Taylor expand the regular potential $V$ about $x'$ to get an expansion in powers of $\varphi_i(t)$ and the increments $(x_i-x_i')$. The first contributions coming from the terms $n=0,1,2$ in the series of expression (\ref{eq.2.exponencial.desarrollada2}) read
\begin{multline}\label{deshea}
{K}(x,x',\beta)=e^{\textstyle-\frac{(x-x')^2}{4\,\beta}}\times
\\
\times\left\lbrace\langle1\rangle-\beta\,V(x')\int_0^1\,dt\,\langle1\rangle-\beta\,\partial_iV(x')
\int_0^1\,dt\, \,\left[(x_i-x'_i)\,t\,\langle1\rangle
+\langle\varphi_i(t)\rangle\right]-\right.
\\
\left.-\frac{\beta}{2}\,\partial_i\partial_jV(x')\int_0^1\,dt\, \,\langle\left[(x_i-x'_i)\,t
+\varphi_i(t)\right]\left[(x_j-x'_j)\,t
+\varphi_j(t)\right]\rangle-\right.
\\
\left.\mbox{}-\ldots+
\frac{\beta^2}{2}\,V^2(x')\int_0^1\int_0^1\,dt\,dt' \langle1\rangle+\ldots\right\rbrace\ .
\end{multline}
Each term of this expansion contains an n-point transition function
\begin{align}\label{npoint}
\nonumber\langle \varphi_{i_1}(t_1)\ldots\varphi_{i_n}(t_n)\rangle:&=
\int\limits_{\substack{\varphi(0)=0\\\varphi(1)=0}} \mathcal{D}\varphi(\tau) \;e^{\textstyle-\int^{1}_0 dt'\,\dot{\varphi}^2(t')/\!4\beta}\,\varphi_{i_1}(t_1)\ldots\varphi_{i_n}(t_n)
\\
&=\left.\frac{\delta}{\delta j_{i_1}(t_1)}\,\ldots\,\frac{\delta}{\delta j_{i_n}(t_n)}
    \ Z[j]\right|_{j\equiv 0}\ .
\end{align}
The generating functional $Z[j]$ is defined as
\begin{align}\label{z}
    \nonumber Z[j]&:=\int\limits_{\substack{\varphi(0)=0\\\varphi(1)=0}} \mathcal{D}\varphi(\tau) \;e^{\textstyle-\int^{1}_0 dt'\,
    \left\{\dot{\varphi}^2(t')/\!4\beta
    +j(t')\varphi(t')\right\}}
    \\
    &=\frac{1}{(4\pi\beta)^{(d+1)/2}}\;e^{\textstyle \beta\int_0^1
    dt\, j(t)(-\partial^2)^{-1}j(t)}\ .
\end{align}
According to expressions (\ref{npoint}) and (\ref{z}), every n-point transition function can be computed in terms of the propagator, which is proportional to the one-dimensional symmetric Green function under the corresponding Dirichlet boundary conditions at $t=0$ and $t=1$
\begin{align}\label{prop}
    \langle
    \varphi_i(t)\varphi_j(t')
    \rangle
    =\delta_{ij}\,\frac{2\beta}{(4\pi\beta)^{(d+1)/2}}\,t(1-t')\qquad {\rm if\ }t<t'\ .
\end{align}

We can now compute the mean values in (\ref{deshea}) and straightforwardly obtain the first terms of an expansion of the heat-kernel in powers of $\beta$ and the increments $\delta_i:=x_i-x'_i$
\begin{align}\label{expxxp}
    \begin{split}
    {K}(x,x',\beta)&=\frac{e^{\textstyle-\frac{\delta^2}{4\,\beta}}}{\left(4\pi\beta\right)^{(d+1)/2}}\ \Biggl\lbrace
    1-\beta \, V
    \mp\frac{\beta}{2}\,\partial_iV\cdot\delta_i+\Biggr.
    \\
    &\hspace{-1.6cm}\mbox{}
    +\frac{\beta^2}{2}\,\left(-\frac{1}{3}\triangle V+V^2\right)-\frac{\beta}{6}\,\partial^2_{ij}V\cdot\delta_i\delta_j
    \pm\frac{\beta^2}{2}\,\left(-\frac{1}{6}\partial_i\triangle V+V\partial_iV\right)\cdot \delta_i\mp
    \\
    &\hspace{-1.6cm}\mbox{}
    \mp\frac{\beta}{24}\,\partial^3_{ijk}V\cdot\delta_i\delta_j\delta_k
    +\frac{\beta^3}{6}\,\left(-\frac{1}{10}\triangle\triangle V+V\triangle V+\frac{1}{2}\partial_iV\partial_iV-V^3\right)+
    \\
    &\hspace{-1.6cm}\mbox{}+\Biggl.
    \frac{\beta^2}{2}\,\left(-\frac{1}{20}\partial^2_{ij}\triangle V+\frac{1}{3}V\partial^2_{ij}V
    +\frac{1}{4}\partial_iV\partial_jV\right)\cdot\delta_i\delta_j-
    \frac{\beta}{120}\,\partial^4_{ijkl}V\cdot\delta_i\delta_j\delta_k\delta_l+\ldots\Biggr\rbrace \ ,
\end{split}\end{align}
where the upper and lower sign apply if the potential and its derivatives are evaluated at the initial point $x'$ or final point $x$, respectively. Notice that, because of the Gaussian prefactor, the integration of the kernel ${K}(x,x',\beta)$ in $x$ implies that each factor $\delta_i$ would give a contribution of order $\beta^{1/2}$. Therefore, the dots in expression (\ref{expxxp}) represent terms of order $\beta^{7/2}$ .

To obtain the asymptotic expansion of the heat-kernel trace we consider the heat-kernel at the diagonal $x=x'$, given by expression (\ref{expxxp}) for $\delta_i=0$
\begin{multline}\label{heaexp}
    {K}(x,x,\beta)=\frac{1}{\left(4\pi\beta\right)^{(d+1)/2}}\ \Biggl\lbrace
    1-\beta \, V
    +\frac{\beta^2}{2}\,\left(-\frac{1}{3}\triangle V+V^2\right)+\Biggr.
    \\\mbox{}\Biggl.
    +\frac{\beta^3}{6}\,\left(-\frac{1}{10}\triangle\triangle V+V\triangle V+\frac{1}{2}\partial_iV\partial_iV-V^3\right)
    +O(\beta^4)\Biggr\rbrace\ ,
\end{multline}
where the potential and its derivatives are now evaluated at $x$. According to eqs.\ (\ref{tra}) and (\ref{traexp}), we integrate expansion (\ref{heaexp}) in $x\in\mathbb{R}^{d+1}$ to obtain the first Seeley-de Witt coefficients (see e.g.\ \cite{Vassilevich:2003xt})
\begin{align}\label{SdW}
b_0&=\int_{\mathbb{R}^{d+1}} 1\\
b_1&=-\int_{\mathbb{R}^{d+1}} V(x)\nonumber\\
b_2&=\frac{1}{6}\int_{\mathbb{R}^{d+1}}\Bigl[-\triangle V\!\left(x\right)+3\,V^2\!\left(x\right)\Bigr]\nonumber\\
b_3&=\frac{1}{60}\int_{\mathbb{R}^{d+1}}
\Bigl[-\triangle\triangle V(x)+
10\,V(x)\,\triangle V(x)+5\,\partial_iV(x)\,\partial_iV(x)-10\,V^3(x)\Bigr]\nonumber\ .
\end{align}
Notice that, as already mentioned, since the base manifold $\mathbb{R}^{d+1}$ has no boundary, the asymptotic expansion of the heat-kernel trace contains only integer powers of $\beta$.

In the next section we will show how this procedure can be generalized to the case where the potential contains a Dirac delta term. In that case, a straightforward application of the method developed in this section would certainly break down at the point of eq.\ (\ref{deshea}) for a Taylor expansion of the potential would lead to products of the Dirac delta and its derivatives evaluated at the same point.

\section{Delta function potential}\label{delpot}

In this section we apply the WF to compute the asymptotic expansion of the heat-kernel trace corresponding to a Schr\"odinger operator $A_\gamma$ whose potential contains a Dirac delta with support on the hyperplane $x_1=0$ of $\mathbb{R}^{d+1}$, i.e.
\begin{align}\label{opedelta}
    A_\gamma=
    -\triangle+V(x)+\gamma\, \delta(x_1)\ ,
\end{align}
where $x\in\mathbb{R}^{d+1}$, $V(x)$ is a regular potential and $\gamma\in\mathbb{R}^+$. We will denote $y\in\mathbb{R}^d$ the coordinates on the surface $x_1=0$.

The heat-kernel at the diagonal corresponding to the operator $A_\gamma$ is given by (see eq.\ (\ref{eq.2.wlf2}))
\begin{align}
\nonumber{K_\gamma}(x,x,\beta)&=
\int\limits_{\substack{x(0)=x\\x(1)=x}} \mathcal{D}x(\tau)
\;e^{\textstyle-\int^{1}_0 dt'\,\left[\dot{x}^2(t')/\!4\beta+\beta\, V\left(x(t')\right)+\beta\,\gamma\,\delta[x_1(t')]\right]}\ .
\end{align}
Notice that we have already rescaled $\tau\rightarrow\beta\tau$. In order to use the results of the preceding section we expand the exponential containing the delta funcion
\begin{multline}\label{pathdelta}
K_\gamma(x,x,\beta)-K_0(x,x,\beta)=\sum_{n=1}^{\infty}\frac{(-\beta\,\gamma)^n}{n!}\int\limits^{1}_0\cdots\int\limits^{1}_0 dt_1 \cdots dt_n\,\times
\\
\times  \int\limits_{\substack{x(0)=x\\x(1)=x}} \mathcal{D}x(\tau)
\;e^{\textstyle-\int^{1}_0 dt'\,\left[\dot{x}^2(t')/\!4\beta+\beta\, V\left(x(t')\right)\right]}\,\delta \left[x_1(t_1)\right]\cdots\,\delta\left[x_1(t_n)\right]\ .
\end{multline}
In this expression we have subtracted the contribution $K_0(x,x,\beta)$ corresponding to the case $\gamma=0$, which has been already computed in the previous section (eq.\ (\ref{heaexp})).

As already mentioned, now we can certainly not expand the delta functions in small fluctuations about the same point; otherwise, divergent products of delta functions would arise. Instead, we regard these path integrals as transition functions under the regular potential $V(x)$ but constrained by the delta functions' terms.

Indeed, the $n$th.\ term in the series of expression (\ref{pathdelta}) receives contributions from all closed ``trajectories'' at the point $x$ with the restriction that at ``times'' $t_1,\ldots t_n$ the ``classical particle'' should ``hit'' the surface $x_1=0$, which is the support of the delta function. Consequently, each path integral with $n$ delta functions is computed as the product of $n+1$ transition functions under the $n$ constraints $x_1(t_1)=\ldots=x_1(t_n)=0$. Expression (\ref{pathdelta}) can be therefore written as
\begin{align}\label{eq.hkreg+delta}
\begin{split}K_\gamma(x,x,\beta)-K_0(x,x,\beta)&=
\\
&\hspace{-3cm}=\sum_{n=1}^{\infty}\left(-\beta\,\gamma\right)^n
\int\limits^{1}_0\cdots\int\limits^{t_3}_0\int\limits^{t_2}_0\,dt_1\,dt_2\,\ldots dt_n
\times
\ \int\limits_{\substack{y(0)=y\\y(1)=y}} \mathcal{D}y(\tau)
\;e^{\textstyle-\int^{1}_{0} dt'\,\dot{y}^2(t')/\!4\beta}
\,\times
\\
&\hspace{-1.5cm}\times\,
\int\limits_{\substack{x_1(0)=x_1\\
x_1(t_{1})=0}} \mathcal{D}x_1(\tau)
\;e^{\textstyle-\int^{t_{1}}_{0} dt'\,\left[\dot{x_1}^2(t')/\!4\beta+\beta\, V\left(x(t')\right)\right]}
\,\times
\\
&\hspace{-1.5cm}\times\,
\prod_{i=1}^{n-1}\int\limits_{\substack{x_1(t_i)=0\\
x_1(t_{i+1})=0}} \mathcal{D}x_1(\tau)
\;e^{\textstyle-\int^{t_{i+1}}_{t_i} dt'\,\left[\dot{x_1}^2(t')/\!4\beta+\beta\, V\left(x(t')\right)\right]}
\,\times\\
&\hspace{-1.5cm}\times\,
\int\limits_{\substack{x_1(t_n)=0\\
x_1(1)=x_1}} \mathcal{D}x_1(\tau)
\;e^{\textstyle-\int^{1}_{t_n} dt'\,\left[\dot{x_1}^2(t')/\!4\beta+\beta\, V\left(x(t')\right)\right]}\ .
\end{split}
\end{align}
In this expression we can see that, for fixed trajectories $y(\tau)$ in the coordinates parallel to the surface, the path integrals in $x_1(\tau)$ are related to one-dimensional heat-kernels in the presence of a regular potential. Notice, however, that the regular potential as a function of the variable $x_1$ depends explicitly on time, due to its dependence on the fixed trajectories $y(\tau)$. In spite of that, these heat-kernels could still be computed along the lines developed in Section \ref{regpot}. Afterwards, this procedure should be repeated on the path integrals on the trajectories $y(\tau)$. Finally, integrating in $x\in\mathbb{R}^{d+1}$ one would obtain the asymptotic expansion of $K(x,x,\beta)$.

Nevertheless, as a matter of technical simplicity, instead of using expression (\ref{eq.hkreg+delta}) we will make further use of this interpretation of the delta potentials in the path integral as constraints on the trajectories. It is more convenient to introduce for each delta function $\delta(x_1(t_i))$ in the path integrals in (\ref{pathdelta}), an extra $d$-dimensional delta function on the coordinates $y$, namely $\delta^{(d)}(y(t_i)-y^{(i)}_0)$. These delta functions are then implemented as the constraint $x(t_i)=(0,y^{(i)}_0)$ as already explained. Next, integration in the variables $y^{(i)}_0$ renders the expected result. Expression (\ref{pathdelta}) then reads
\begin{multline}\label{eq.hkreg+delta2}
K_\gamma(x,x,\beta)-K_0(x,x,\beta)=
\\
\hspace{-3cm}=\sum_{n=1}^{\infty}\left(-\beta\,\gamma\right)^n
\int\limits^{1}_0\cdots\int\limits^{t_3}_0\int\limits^{t_2}_0\,dt_1\,dt_2\,\ldots dt_n\,\times
\int_{\mathbb{R}^d}\ldots \int_{\mathbb{R}^d}dy^{(1)}\ldots dy^{(n)}
\,\times
\\
\times\,
K(x^{(1)},x,\beta t_1)\,
\times\,
\ldots\,
K(x^{(i+1)},x^{(i)},\beta(t_{i+1}-t_i))\,
\ldots\,\times\,
K(x,x^{(n)},\beta-\beta t_n)\ ,
\end{multline}
where $x^{(i)}=(0,y^{(i)})$. The asymptotic expansions of the kernels $K$ in expression (\ref{eq.hkreg+delta2}) can be read from expression (\ref{expxxp}).

A remark is now in order. It could be argued that if the integrals in the coordinates $y^{(i)}$ were absent and the coordinates $y^{(i)}$ took a fixed value, say $y^{(i)}=0$, then expression (\ref{eq.hkreg+delta2}) would give the transition function in the presence of a $d+1$-dimensional delta with support at the origin. However, it is well-known that this kind of potential gives an ill-defined Schr\"odinger Hamiltonian. This is consistent with the fact that if the integrals in the $y^{(i)}$ were absent in expression (\ref{eq.hkreg+delta2}), then the subsequent integrals in $t_i$ would be divergent.

Turning back to expression (\ref{eq.hkreg+delta2}), after replacing the heat-kernels by their expansions given in (\ref{expxxp}) we can finally integrate in $x\in\mathbb{R}^{d+1}$ to obtain the small $\beta$-asymptotic expansion of the difference between the heat-kernel traces
\begin{multline}\label{headel}
{\rm Tr}\,e^{-\beta A_\gamma}-{\rm Tr}\,e^{-\beta A_0}=\frac{\gamma}{(4\pi\beta)^{(d+1)/2}}\int\,d^{d}y\,
\Biggl\{
-\beta+\frac{\sqrt{\pi}}{4}\,\gamma\,\beta^{3/2}+\left(V-\frac{1}{6}\,\gamma^2\right)\beta^2\Biggr.-\\
\Biggl.-\frac{\sqrt{\pi}}{32}\,\gamma\left(8V-\gamma^2\right)\beta^{5/2}+
\frac{1}{6}\left(\triangle V-3V^2+\gamma^2\,V\right)\beta^3+O(\beta^{7/2})
\Biggr\}\ ,
\end{multline}
where $A_0$ denotes the regular Schr\"odinger operator given by (\ref{opedelta}) for $\gamma=0$, whose heat-kernel expansion was computed in Section \ref{regpot}.

From expansion (\ref{headel}) we read the corrections $\Delta b_n$ to the first Seeley-de Witt coefficients $b_n$ for $n=1,2,3$ (see eqs.\ (\ref{SdW})) due to the Dirac delta superimposed to the regular potential $V$. Moreover, we see that the presence of the delta potential implies the appearance of half-integer powers of $\beta$ in the heat-kernel expansion, as is the general case when the base manifold has boundaries.

We display the corrections to the first Seeley-de Witt coefficients, as well as the first coefficients $\Delta b_{n/2}$, with odd $n$, corresponding to half-integer powers of $\beta$:
\begin{align}\label{eq.5.coef.delta}
\begin{split}
\Delta b_{1}&=-\gamma
\\
\Delta b_{3/2}&=\frac{\sqrt{\pi}}{4}\,\gamma^2
\\
\Delta b_{2}&=\gamma\left(V-\frac{1}{6}\,\gamma^2\right)
\\
\Delta b_{5/2}&=-\frac{\sqrt{\pi}}{32}\,\gamma^2\left(8V-\gamma^2\right)\\
\Delta b_3&=\frac{1}{6}\,\gamma\left(\triangle V-3V^2+\gamma^2\,V\right)\ .
\end{split}
\end{align}
The first four coefficients displayed in (\ref{eq.5.coef.delta}) can be obtained as a particular case of the results in \cite{Bordag:1999ed}.

\section{Casimir force between two semitransparent boundaries}\label{casfor}

The calculations of Section \ref{delpot} can also be carried out in the presence of Dirac delta functions with support on several hyperplanes. As mentioned in the introduction, a Dirac delta can be used to model the boundary conditions of the TE modes of the electromagnetic field in the vicinity of plasma layers for small values of the fluctuations of the charge density and current in the layer. Consequently, as an application, we will compute in this section the Casimir force between two parallel plasma layers due to the quantum oscillations of a scalar field which is weakly coupled to the Dirac deltas. This has already been computed for a general coupling by using an explicit expression of the corresponding Green function in \cite{Bordag:1992cm}.

We consider the following Euclidean action for a massive scalar field $\phi(x)$ defined on $x\in\mathbb{R}^{d+1}$
\begin{align}\label{act1}
    S[\phi]=\frac{1}{2}\int_{\mathbb{R}^{d+1}}\partial_i \phi(x)\partial_i\phi(x)+
    m^2\phi^2(x)+\gamma\left[\delta(x_1+L/2)+\delta(x_1-L/2)\right]\phi^2(x)\ ,
\end{align}
where the interaction with the Dirac delta functions imposes semitransparent boundary conditions on the scalar field at the hyperplanes $x_1=-L/2$ and $x_1=L/2$. Notice that the $d+1$ coordinates in Euclidean space consist in the Euclidean time, the spatial coordinate $x_1$, describing the position of the layers, and $d-1$ spatial coordinates parallel to the layers.

The corresponding heat-kernel at the diagonal is given by (see eq.\ (\ref{eq.2.wlf2}))
\begin{multline}
{K}(x,x,\beta)=\frac{1}{\textstyle{(4\pi\beta)^{d/2}}}\ e^{-\beta m^2}\,\times\\
\times\int\limits_{\substack{x_1(0)=x_1\\x_1(1)=x_1}}\mathcal{D}x_1(\tau)
\;e^{\textstyle-\int^{1}_0 dt'\,\left[\dot{x_1}^2(t')/\!4\beta+\beta\, \gamma\,\delta \left[x_1(t')+L/2\right]+\beta\, \gamma\,\delta \left[x_1(t')-L/2\right]\right]}\ ,
\end{multline}
where we have factorized the contributions from Euclidean time and the $d-1$ coordinates parallel to the layers. As we have  mentioned, we are interested in the Casimir energy in the weak coupling regime,i.e.\ for small values of the coupling parameter $\gamma$.

A rescaling of the dimensionful constants $m$ and $\gamma$ in terms of the distance $L$ between the layers leads us to the following expression for $\gamma L\ll 1$
\begin{align}\label{eq.4.exponencial.desarrollada}
\begin{split}
{K}(x,x,\beta)&=\frac{1}{\textstyle{(4\pi\beta)^{d/2}}}\ e^{-\beta m^2}
\int\limits_{\substack{x_1(0)=x_1\\
x_1(1)=x_1}} \mathcal{D}x_1(\tau)
\;e^{\textstyle-\int^{1}_0 dt'\,\dot{x_1}^2(t')/\!4\beta}\;\times
\\
&\hspace{1cm}\times\left\lbrace1-\beta\,\gamma\int^1_0 dt\, \delta\left[x_1(t)+L/2\right]-
\beta\,\gamma \int^1_0 dt\, \delta\left[x_1(t)-L/2\right]+\right.
\\
&\hspace{3.cm}+\left.\mbox{}\beta^2\,\gamma^2 \int^1_0\int^1_0 ds\, dt\, \delta\left[x_1(s)+L/2\right]\,\delta\left[x_1(t)-L/2\right]+\ldots\right\rbrace\ .
\end{split}
\end{align}
The first term in the series (\ref{eq.4.exponencial.desarrollada}) is related to the divergent energy of empty space, while the second and third terms contribute to the self-energy of each layer, which is easily shown to be independent of the separation $L$ between the layers. It is the fourth term that gives the leading contribution in $\gamma$ to the interaction energy between the layers. This term, which we denote by $K_{L}(x,x,\beta)$, can be computed as described in Section \ref{delpot} and gives the following contribution to the heat-kernel trace
\begin{align}\label{eq.4.pre.heat.kernel}
\begin{split}
&\int_{\mathbb{R}^{{d+1}}}dx\,K_{L}(x,x,\beta):=
\\
&=\frac{\textstyle{T\,A_{d-1}}}{\textstyle{(4\pi\beta)^{d/2}}}\ e^{-\beta m^2}\,\beta^2\,\gamma^2
\,\times
\\
&\times\,\int_{\mathbb{R}}dx_1\,\int\limits_{\substack{x_1(0)=x_1\\x_1(1)=x_1}} \mathcal{D}x_1(\tau)
\;e^{-\int^{1}_0 dt'\,\dot{x_1}^2(t')/\!4\beta}\;
\int^1_0\int^1_0 ds\, dt\, \delta\left[x_1(s)+L/2\right]\,\delta\left[x_1(t)-L/2\right]
\\
&=\frac{\textstyle{T\,A_{d-1}}}{\textstyle{(4\,\pi\,\beta)^{d/2}}}\ e^{-\beta m^2}2\beta^2\,\gamma^2\,\times
\\
&\times
\,\int_{\mathbb{R}}dx_1\,
\int\limits^{1}_0\int\limits^{t}_0 ds\,dt\ \frac{\textstyle{e^{-(x_1+L/2)^2/\!4\beta s}}}{\textstyle{\sqrt{4\,\pi\,\beta\,s}}}\,\frac{\textstyle{e^{-L^2/\!4\beta (t-s)}}}{\textstyle{\sqrt{4\,\pi\,\beta\,(t-s)}}}\,\frac{\textstyle{e^{-(x_1-L/2)^2/\!4\beta (1-t)}}}{\textstyle{\sqrt{4\,\pi\,\beta\,(1-t)}}}
\\
&=\frac{\textstyle{T\,A_{d-1}}}{\textstyle{(4\,\pi\,\beta)^{d/2}}}\,\frac{\textstyle{\beta\,\gamma^2}}{\textstyle{4}}
\,e^{-\beta\,m^2}\,\text{erfc}(L/\sqrt{\beta})\ .
\end{split}
\end{align}
In this expression $T$ represents the infinite length of the time interval, $A_{d-1}$ is the volume of the $(d-1)$-dimensional layers and $\text{erfc}$ is the complementary error function.

The Casimir energy $E_0$ per unit area of the layers can be defined by means of the 1-loop correction $\Gamma^{(1)}_{\rm eff}$ to the effective action $\Gamma[\phi]$ (given by the second term in eq.\ (\ref{effact})) as $E_0:=\Gamma^{(1)}_{\rm eff}/T\,A_{d-1}$. Therefore, the interaction energy per unit area between the layers due to the vacuum oscillations of the scalar field reads (see eq.\ (\ref{dethea}))
\begin{equation}\label{enac}
    E_0=-\frac{1}{T\,A_{d-1}}\frac{1}{2}\int_0^\infty\frac{d\beta}{\beta}
    \int_{\mathbb{R}^{{d+1}}}dx\,K_{L}(x,x,\beta)\ ,
\end{equation}
whereas the Casimir pressure on the layers can be computed as
\begin{equation}\label{press}
    p:=-\frac{dE_0}{dL}\ .
\end{equation}
Replacing the leading contribution of the interaction between the layers to the heat-kernel given by eq.\ (\ref{eq.4.pre.heat.kernel}) into eqs.\ (\ref{enac}) and (\ref{press}) we obtain the pressure in the weak coupling regime $\gamma L\ll 1$
\begin{equation}\label{pressweak}
    p=-\frac{\gamma^2}{(4\pi)^{(d+1)/2}}\,(m/L)^{d/2-1/2}\,K_{d/2-1/2}(2mL)\ ,
\end{equation}
in terms of the modified Bessel function.

Let us finally mention two limiting cases of expression (\ref{pressweak}). Firstly, we consider the Casimir pressure for large separation $L$ of the layers. If we take the limit $mL\gg 1$ in eq.\ (\ref{pressweak}) we obtain
\begin{equation}\label{press2}
    p\simeq-\frac{\gamma^2}{2^{d+2}\pi^{d/2}}\,\frac{m^{d/2-1}}{L^{d/2}}\,e^{-2mL}\ .
\end{equation}
As remarked in \cite{Bordag:1992cm}, the exponential decrease with the product $mL$ of the Casimir attraction due to the vacuum oscillation of massive fields justifies the consideration of photons for the determination of the Casimir effect.

Secondly, we consider the massless limit of expression (\ref{press}). For $mL\ll 1$ we obtain
\begin{equation}\label{press3}
    p\simeq
    \left\{
    \begin{array}{lcr}
    \frac{\textstyle\gamma^2}{\textstyle4\pi}\,\log{(Lm)}&&{\rm if\ }d=1\ ,\\\mbox{}\\
    -\frac{\textstyle\gamma^2}{\textstyle(4\pi)^{(d+1)/2}}
    \,\frac{\textstyle\Gamma((d-1)/2)}{\textstyle2}\,\frac{\textstyle1}{\textstyle L^{d-1}}&&{\rm if\ }d\geq 2\ .
    \end{array}
    \right.
\end{equation}
The first line of eq.\ (\ref{press3}) is consistent with the fact, pointed out in \cite{Milton:2004vy}, that in $1+1$ dimensions ($d=1$) the massless field generates a Casimir pressure which is not analytic in $\gamma$ for small values of this parameter. For $d=3$, the second line can be obtained as the weak coupling limit of the result in \cite{Bordag:2005by}.

\section{Conclusions}

In this article we have considered the generalization of the Worldline Formalism' (WF) techniques to study scalar fields under semi-transparent boundary conditions. In particular, we determine the asymptotic expansion for small values of the proper time of the heat-kernel corresponding to a Schr\"odinger operator whose potential contains Dirac delta functions with support on flat surfaces.

In Section \ref{delpot} we have studied the case in which the potential contains a regular term superimposed to a single delta function. In Section \ref{casfor} we have considered two delta functions with support on parallel surfaces and we have computed the Casimir energy due to the vacuum oscillations of a scalar field weakly coupled to the delta functions.

We have shown that the WF, which has been extensively used as a very efficient and intuitive tool for 1-loop calculations, is also very suitable for handling with delta functions. Indeed, some advantages of this formalism are still present in the case of a potential with delta functions. On the one hand, although we have computed the small $\beta$-asymptotic expansion of the heat-kernel trace, the WF provides a mechanism for computing the heat-kernel $K(x,x',\beta)$ out of the diagonal, i.e.\ for $x\neq x'$; this determines, e.g., the expansion of the propagator at small distances.

On the other hand, the techniques of the WF do not require the knowledge of the explicit solutions of the corresponding Schr\"odinger operator. For this reason, we were able to compute the effects of the delta function on the Seeley-de Witt coefficients corresponding to a regular potential. In relation to the aforementioned application of this model to the study of the Casimir energy of the TE modes in the vicinity of flat plasma layers, the calculation of the present article would be related to the case where the space between the layers were filled by a dielectric with position dependent-permittivity. A natural generalization of our result, would be to consider the case in which the Dirac delta functions have support on curved surfaces.

\section*{Acknowledgements}

The authors thank Prof.\ H.\ Falomir for his helpful contributions at every stage of this work. This work was partially supported by CONICET (PIP 01787), ANPCyT (PICT 00909) and UNLP (Proy.~11/X492), Argentina.

\end{document}